# Quadruply Bonded $Mo_2$ Molecules: An Innate Emitter-Resonator Quantum System in Free Space

Miao Meng, Ying Ning Tan, Zi Cong He, Zi Hao Zhong, Jia Zhou, Yu Li Zhou, Guang Yuan Zhu, Chun Y. Liu*

Department of Chemistry, College of Chemistry and Materials Science, Jinan University, 601 Huang–Pu Avenue West, Guangzhou 510632, China

* Correspondence: tcyliu@jnu.edu.cn

*Dedicated to the discovery of metal–metal quadruple bonds by F. Albert Cotton 60 years ago*

**Abstract**

In recent decades, significant progress has been made in constructing and studying individual quantum systems based on two-level atoms (molecules) and photons. Here we demonstrate that the quadruply bonded $Mo_2$ unit, with a Mo–Mo bond distance as short as 2.1 Å, functions as an innate emitter–resonator molecular quantum system capable of trapping visible-light photons between the two molybdenum atoms under ambient conditions, thereby generating an intense quantized local electromagnetic field with an extremely small mode volume. The resonance fluorescence spectra of three $Mo_2$ complexes indicate that the intermetallic Mo→Mo charge-transfer transition is coherently coupled to the local scattering field, exhibiting vacuum Rabi splitting and Mollow triplets. Resonant coupling of single molecules and *N*-molecule ensembles to the scattered light through sideband excitation produces a sequence of discrete optical modes spanning a broad wavelength range, with polaritonic transitions identical to those observed in $Ni_2$-based systems. These results establish the $Mo_2$ molecule as an independent emitter–resonator integrated quantum system that enables quantum-optical experiments in free space using conventional spectroscopic instrumentation. This work extends quantum electrodynamics into molecular science, providing new insights into metal–metal bonding, molecular physics, and light–matter interactions.

## 1. Introduction

Traditionally, strong coherent coupling between matter and light at the level of individual quantum system is realized using optical cavities, such as Fabry–Pérot resonators composed of highly reflective mirrors or layers (Figure 1A-a), into which atoms[1,2] and photons are introduced under carefully controlled conditions. Such cavity quantum electrodynamics (C-QED) experiments have enabled the observation of fundamental quantum phenomena, including vacuum Rabi splitting[3] and field quantization,[2] but typically rely on cryogenic temperatures, high-Q resonators, and sophisticated experimental infrastructures. For example, trapped-ion platforms employ laser cooling to isolate individual ions at ultralow temperatures,[1,4,5,6] while photon-box experiments use highly excited Rydberg atoms interacting with single microwave photons in superconducting cavities.[1,2,7] Within these controlled environments, atomic transitions can interact coherently with confined electromagnetic (EM) fields, forming hybrid light–matter states known as exciton-polaritons.[1] The eigenstates of such systems are commonly described by the Jaynes–Cummings (JC) Hamiltonian, which models the resonant interaction between a two-level emitter and a single photonic mode at the single-quantum level.[8,9,10] Extensions of this framework to multiple emitters, such as the Tavis–Cummings model, reveal collective quantum effects arising from ensemble coupling.[11]

Over the past two decades, significant progress has also been made in achieving strong light–matter coupling at the single-molecule level.[12, 13, 14] By focusing laser excitation onto isolated molecules embedded in condensed-phase environments, individual molecular quantum systems with small effective mode volumes have been realized.[15,16] Single-molecule spectroscopy under these conditions has revealed a range of quantum-optical phenomena, including photon antibunching,[17,18] squeezed-light emission,[19,20] vacuum Rabi splitting,[13] and Mollow triplets,[13,21,22] thereby establishing molecules as viable platforms for exploring quantum optics. These advances have bridged quantum optics with chemistry and materials science, providing a new methodology for manipulating chemical reactivity and energy transfer using quantized electromagnetic fields.[23]

A central strategy for enhancing light–matter interactions involves reducing the photonic mode volume $V$ to increase the local field amplitude. Plasmonic nanocavities formed by metallic nanoparticles separated by sub-nanometer gaps have been extensively explored for this purpose.[24,25] In the nanoparticle-on-mirror geometry and related setups, such as the tip-to-tip configuration shown in Figure 1A-b, mode volumes approaching $10^{-2}\ \mathrm{nm}^3$ have been achieved.[25] The enhanced local EM field is capable of driving strong optical transitions in the visible range.[24] When the interparticle separation crosses over the critical distance ($d_{\mathrm{QR}}$), for example, $d_{\mathrm{QR}} \sim$ 0.31 nm for gold dimers,[26] the system enters the quantum tunneling regime, in which charge-transfer plasmons (PCT) emerge due to electron tunneling between the surfaces of metal particles.[27] These advances motivate our exploration of molecular analogues in which an atomic-scale resonator is intrinsically built into the molecular system. Dinuclear transition-metal complexes that feature both metal–metal bonding[28] and non-bonding[29] structures have been naturally selected to demonstrate this concept. In such systems, the diatomic metal core ($M_2$) acts as an atomistic optical resonator, producing an enhanced local EM field through interatomic charge-transfer processes. We recently demonstrated that a dinickel ($Ni_2$) complex functions as an integrated emitter–resonator quantum system with an effective mode volume $V \approx 10^{-3}\ \mathrm{nm}^3$ (Figure 1A-c), in which metal–metal charge-transfer transitions are coherently coupled to the scattering field.[29] The $Ni_2$ system exhibits hallmark Jaynes–Cummings (JC) physics under ambient conditions, including vacuum Rabi splitting, Mollow triplets, and an unconventional collective coupling scaling $\Omega_{\mathrm{N}}$ proportional to $N\sqrt{N}\,\Omega_1$ which differs fundamentally from the $\sqrt{N}\,\Omega_1$ scaling predicted by the Tavis–Cummings model.[11, 29]

Dimolybdenum ($Mo_2$) complexes provide an especially compelling platform for testing the generality of this molecular quantum-optical framework[28] and the quantum limit of the dimetal configuration.[26,27] Since the discovery of metal–metal quadruple bonds by Cotton and co-workers,[30] quadruply bonded $Mo_2$ units have been recognized as prototypical systems in metal–metal bond chemistry.[31] The electronic configuration ($\sigma^2\pi^4\delta^2$) for the $Mo_2$ unit gives rise to a short metal–metal separation of approximately 2 Å (Figure 1A-d) and a characteristic electric-dipole-allowed $\delta \rightarrow \delta^*$ transition.[32,33,34,35,36] Importantly, the Mo–Mo distance in these complexes places the system well within the quantum tunneling regime, rendering the $Mo_2$ unit an ideal molecular analogue of a contact-scale plasmonic cavity ($d$ < 0).[26,27] In contrast to the $Ni_2$ complex systems, in which the $Ni_2$ unit consists of two unbonded $Ni^{2+}$coordination moieties, the quadruply bonded $Mo_2$ core introduces a markedly different electronic structure while preserving the essential diatomic geometry. This distinction enables a stringent assessment of whether the observed quantum-optical phenomena originate from specific electronic configurations or from the intrinsic resonator character of the diatomic metal unit itself. Elucidating the nature of light–matter coupling in $Mo_2$ complexes therefore not only advances our knowledge of metal–metal bond chemistry, but also provides a critical validation of the universality of molecular Jaynes–Cummings physics.

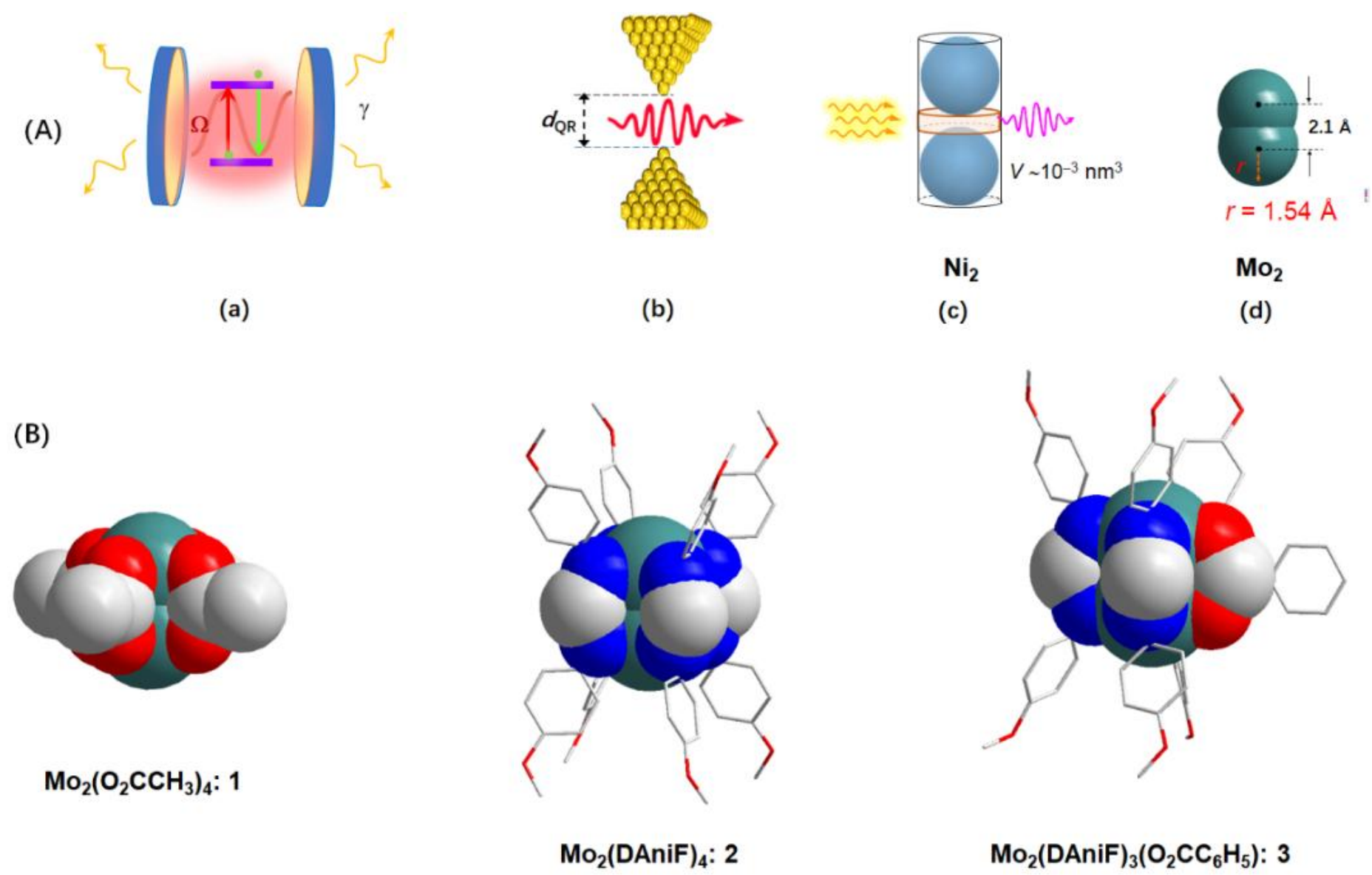

**Figure 1. Illustration of optical cavity minimization across different length scales and the molecular structures of the $Mo_2$ complexes studied**. (A) Schematic representations of (a) a high-Q Fabry–Pérot optical cavity, (b) a sub-nanometer plasmonic cavity formed by

metallic nanostructures in the quantum tunneling regime characterized by a critical distance $d_{QR}$, (c) a $Ni_2$ diatomic optical resonator, and (d) a quadruply-bonded $Mo_2$ diatomic optical resonator with two conductively coupled Mo atoms ($d < 0$). (B) Space-filling models of $Mo_2(O_2CCH_3)_4$ (**1**), $Mo_2(DAniF)_4$ (**2**), and $Mo_2(DAniF)_3(O_2CC_6H_5)$ (**3**), illustrating the common $Mo_2$ core and differing ligand environments.

In this work, we investigate three $Mo_2$ complexes—$Mo_2(O_2CCH_3)_4$ (**1**), $Mo_2(DAniF)_4$ (**2**) (DAniF = *N*, *N*′-di(p-anisyl)formamidinate), and $Mo_2(DAniF)_3(O_2CC_6H_5)$ (**3**) (Figure 1B)—that share a common quadruply bonded $Mo_2$ core but differ in their ligand environments. Despite substantial differences in electronic structure and ligand-based electronic transitions, all three complexes exhibit remarkably similar quantum-optical responses. We show that the $Mo_2$ unit traps visible-light photons between the two molybdenum atoms, and that radiative relaxation of the excited states generates a quantized local scattering field composed of discrete modes. Resonance fluorescence measurements reveal vacuum Rabi splitting, Mollow triplets, and collective coupling behavior that closely mirror the observations in $Ni_2(DAniF)_4$.[29] These results establish quadruply bonded $Mo_2$ molecules as independent, innate emitter–resonator quantum systems operating in free space under ambient conditions, thereby generalizing molecular quantum optics beyond a single non-bonding $M_2$ motif.

## 2. Results and Discussion

### 2.1 Photon scattering by the $Mo_2$ unit in the complex (1) upon non-resonant excitation

$Mo_2(O_2CCH_3)_4$ (**1**) exhibits a single broad fluorescence band with a maximum at around 400 nm when excited by the second-harmonic generation (SHG) from a fundamental laser beam with excitation wavelengths ($\lambda_{ex}$) between 560 and 600 nm (Figure 2A). In particular, this emission is distinct from the $\delta^* \rightarrow \delta$ relaxation of the quadruply bonded $Mo_2$ core, which occurs at lower energy of approximately 450 nm,[30,32,36] and thus, cannot be assigned to conventional Stokes fluorescence. Similar featureless emission bands have previously been observed in the $Cu_2$ (420 nm)[28] and $Ni_2$ (380 nm)[29] molecular systems in which the two metal atoms are not covalently bonded. The similar fluorescence spectra for different dimetal units under excitation of varying laser wavelengths indicate that the emissions are not governed by the electronic states. Therefore, this fluorescence feature for the $M_2$ (M = Cu, Ni, Mo) complexes has been attributed to the averaged photon scattering associated with intermetallic charge-transfer processes.[29] These results confirm that the emissions in the 400 nm region result from the radiative decay of the $M_2$ unit excited by incident light ($\omega_{in}$), [29] producing the $CT_0$ scattering mode ($\omega_{Mo2} = \omega_L$) (Figure 2B). This local scattering EM field can be extremely intense due to the atomistic confinement, analogous to PCT modes within the metallic junction.[24] Notably, the emission energies around 400 nm for these $M_2$ cores closely matches the high-order PCT transitions (~ 3 eV) theoretically predicted for metallic nanoparticle dimers in the quantum contact regime ($d < 0$). [27] The transition energy for $Mo_2$ lies between the corresponding values observed for the $Ni_2$ and $Cu_2$ systems.[29,28] The blue shift of the emission spectrum for $Mo_2$ with $d_{Mo\text{-}Mo} = 0.04$ nm, relative to the 420 nm emission for $Cu_2$ with $d_{Cu\text{-}Cu} = 0.12$ nm, represents a conductive coupling feature for systems with $d < 0$.[26] These results demonstrate that like the nonbonding $Ni_2$ system,[29] the quadruply bonded $Mo_2$ unit acts as an atomistic optical resonator, enabling conversion of classical incident light into a strong local scattering field under ambient conditions. At the single-molecule level, the scattered light naturally exhibits non-classical optical features, such as photon antibunching and squeezed states.[16,17,19]

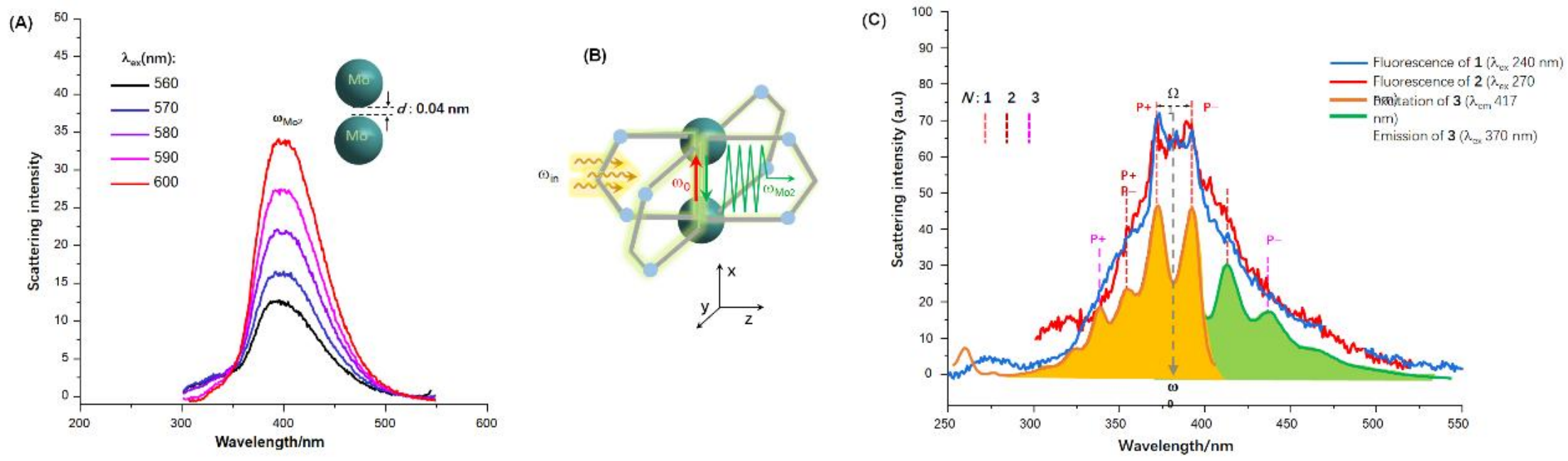

**Figure 2. Optical responses of the $Mo_2$ complex molecules.** (A) Fluorescence spectra of $Mo_2(O_2CCH_3)_4$ (**1**) under incoherent excitation, showing the thermally averaged scattering

continuum associated with the $Mo_2$ unit. (B) Schematic illustration of the conversion of incident classical light ($\omega_{in}$) into quantized scattering light ($\omega_{Mo2}$) mediated by intermetallic charge transfer between the two Mo atoms. (C) Resonance fluorescence spectra of the $Mo_2$ complexes, displaying Rabi doublets for single molecules ($N$ = 1) and collective $N$-molecule ensembles (N = 2 and 3), shown in comparison with the corresponding excitation–emission spectra.

### 2.2 Resonance fluorescence spectra for the strongly coupled $Mo_2$ complexes

Excitation of complexes **1** and **2** with monochromatic ultraviolet light in the range of $\lambda_{ex}$ = 240–270 nm produces characteristic resonance fluorescence spectra featuring pairs of emission peaks symmetrically distributed about the 382 nm position, as shown in Figure 2C. This spectral profile closely resembles that observed for the $Ni_2$ analogue,[29] thus, representing the resonance fluorescence for the strongly coupled $Mo_2$ system, despite their distinct electronic structures. Accordingly, the two dominant peaks at 372 and 392 nm, centered at the resonance frequency $\omega_0$ = 382 nm, are assigned to the upper (P+) and lower (P − ) exciton–polaritonic branches of the dressed single JC molecules.[29] Vacuum Rabi splitting of the resonance gives a coupling strength $\Omega_1$ = 1380 (±10) $cm^{-1}$, identical to that measured for $Ni_2$.[29] The close correspondence in both peak positions and splitting scale provides compelling evidence that the resonance fluorescence observed for the $Mo_2$ complexes originates from strong coupling between the intermetallic charge-transfer excitation ($CT_0$) and the local scattering field at the resonance ($\omega_0$). The nearly identical fluorescence lineshapes for $Mo_2$ and $Ni_2$ further indicate that both dimetal systems share a common physical origin for photon emission and a same normal-mode frequency $\omega_0$, corresponding to interatomic charge transfer via quantum tunneling. This strong resonant coupling under weak, non-resonant excitation is remarkable. By analogy with the $Ni_2$ system,[29] the $Mo_2$ molecules are dressed by photon antibunching at resonance ($\omega_L \approx \omega_0$). [17,18,19, 25]

In addition to the single-molecule polaritonic doublet, additional emission bands appear at approximately 355 and 413 nm in the spectra (Figures 2C and 3A-3C), corresponding to the P+ and P − Rabi branches of the collectively coupled two-molecule ($N = 2$) JC system, as previously observed for $Ni_2$.[29] The Rabi splitting states of the three-molecule ensemble ($N = 3$) are confirmed by the combined excitation–emission spectra of complex **3** (Figure 2C, Table S1).[37,38] Specifically, the absorption at 336 nm ($P^+$) and emission at 442 nm ($P^-$) are in quantitative agreement with the polaritonic transitions previously reported for $Ni_2$.[29] The corresponding collective coupling strengths, $\Omega_2 = 4430$ cm$^{-1}$ ($N = 2$) and $\Omega_3 = 7140$ cm$^{-1}$ ($N = 3$) follow the unconventional $N\sqrt{N}\Omega_1$ ($\Omega_1 = 1380$ cm$^{-1}$) scaling for collective coupling in the $M_2$ molecular JC system.[29] For $N = 3$, the normalized coupling strength reaches $\eta = \Omega/(2\omega_0) \approx 0.14$, placing the $Mo_2$ system firmly in the ultrastrong-coupling regime ($\eta > 0.1$). In this collective coupling framework,[29] the factor $N$ arises from the simultaneous excitation of $N$ molecular dipoles, yielding a total dipole moment $N\mu$, while the $\sqrt{N}$ factor reflects the absorption of $N$ photons by the dressed $N$-molecule ensemble, highlighting the intrinsically nonlinear optical response of the JC molecular system.[8,9,10] This behavior contrasts sharply with the $\sqrt{N}\Omega_1$ scaling predicted by the Tavis–Cummings model,[11] where collectively coupling involves a huge number of identical molecules, and strong dipole–dipole interactions suppress full excitation of Dicke states.[1,39] The present results, consistent with theoretical predictions, suggest that squeezed scattering fields generated by the JC molecule relax the dipole blockade constraints,[28,40,41] thereby enabling ultrastrong collective coupling to emerge readily in the $M_2$ system.

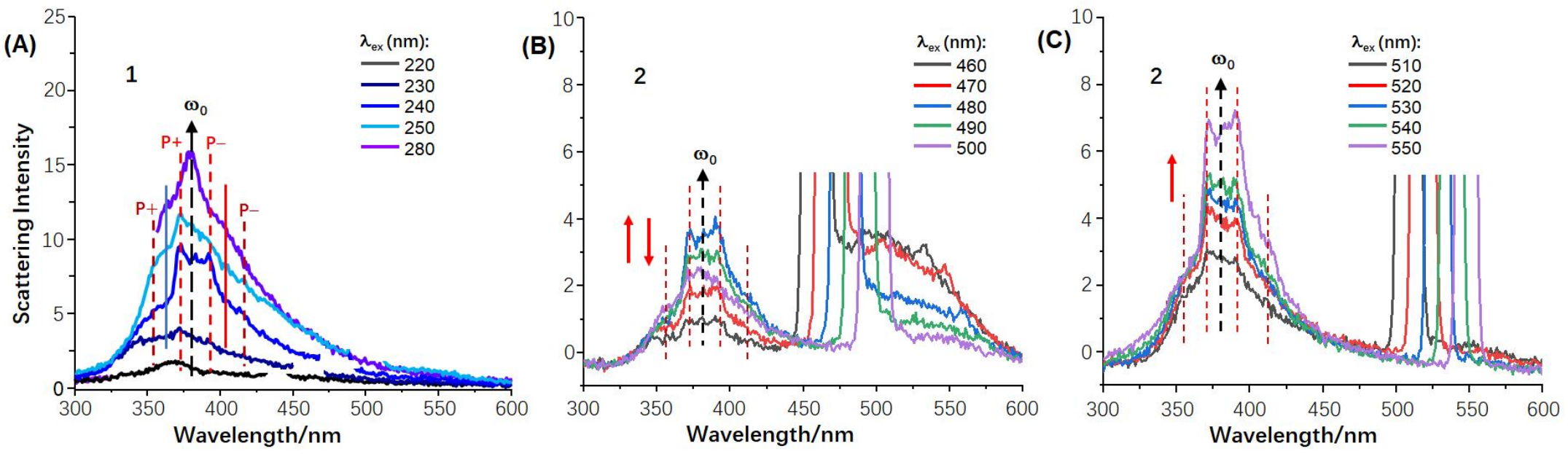

**Figure 3. Resonance fluorescence spectra of the strongly coupled $Mo_2$**

**Jaynes–Cummings molecular systems.** Resonance fluorescence spectra for $Mo_2$ complexes **1** and **2** under weak far-off-resonant excitation conditions. (A) Resonance fluorescence of complex **1** showing the Mollow triplets (solid lines) at $\omega_0$ and P+ for $N$ =1 (372 nm), together with Rabi-split polaritonic doublets (dashed lines) for $N$ = 1 (red) and $N$ = 2 (dark red) under high-energy, non-resonant excitation. The blue and red sidebands of the Mollow triplet at $\omega_0$ are marked using blue and red solid lines, respectively. (B, C) Representative resonance fluorescence spectra of complex **2** excited via second-harmonic generation (SHG), with the fundamental laser wavelength varied from 460 to 550 nm. The fluorescence intensity exhibits pronounced fluctuations as a function of excitation wavelength, as indicated by the red arrows, reflecting excitation-dependent polaritonic coherence.

Under far-off-resonant excitation, the dressed molecule **1** exhibits resonance fluorescence characterized by the Rabi components P+ and P − for the single-molecule system, together with a Mollow triplet at the resonance frequency $\omega_0$,[29] with two sidebands displaced by $\omega_0 \pm \Omega_1'$, where $\Omega_1' = 1450$ cm$^{-1}$ (Figure 3A, Table S2). The spectra with two shoulder peaks at 356 nm (P+) and 413 nm (P−) further reveal vacuum Rabi splitting of the resonance under simultaneous excitation of two molecules, yielding a collective coupling strength of 3880 cm$^{-1}$, comparable to that observed for $Ni_2$ ($N$ = 2).[29] In addition, a distinct triplet structure appears at 372 nm, a position corresponding to the upper Rabi branch (P+) of $N$ = 1. This feature represents a distorted Mollow triplet centered at $\omega_{372}$ with two asymmetric satellites at $\omega_{372} \pm \Omega_1'$, which therefore results from polariton excitation by photons scattered at the $\omega_{372}$ sideband[42] under external excitation at 230 and 250 nm (Figure 3A). These observations reveal the differences in resonant coupling dynamics between **1** the $Ni_2$ analogue.[29] For complex **2**, the characteristic resonance fluorescence observed under excitation at $\lambda_{ex}$ = 270 nm (Figure 2C) is also reproduced under excitation of the second-harmonic generation (SHG) in a broad wavelength range, confirming the resonance Rabi splitting arising from both single-molecule and collective coupling.

As shown in Figures 3B and 3C, the fluorescence spectra exhibit symmetrically distributed emission pairs about $\omega_0$, with peak positions corresponding closely to the polaritonic transitions predicted for the Rabi splitting states of the $N$ = 1 and $N$ = 2 systems (Table S2).[29] It appears that the SHG excitations, with fundamental wavelengths ranging from 460 to 550 nm, provides a more favorable phase-matching environment for generating doubled-frequency light in the 230–275 nm range, enabling efficient internal resonant coupling under non-resonant external excitation. Notably, mapping the SHG excitation wavelength from 460 to 550 nm reveals pronounced fluctuations in the resonance fluorescence intensity (Figures 3B and 3C), suggesting polaritonic coherence in the $Mo_2$ molecular system.

### 3.3 Transformation from the Rabi splitting states to the Mollow triplet states

When excited with monochromatic laser light in the 320–340 nm range, complex **1** in a dilute dichloromethane solution exhibits an intense triplet fluorescence profile centered at 392 nm, corresponding to the lower polaritonic branch (P − ) for the driven single molecules (Figure 4A, Table S2). Two sidebands are symmetrically displaced from the central line by $\pm \Omega_1'$, forming a Mollow triplet at $\omega_{392}$, consistent with the spectra of **2**.[28] This triplet is therefore assigned to resonance fluorescence arising from excitation of the P − polariton by the scattering light of $\omega_{392}$ mode.[42] The spectral profile deviates from the ideal Mollow structure, exhibiting asymmetry and partial distortion, which can be attributed to incoherent excitation and to modifications imposed by the squeezed vacuum field intrinsic to the JC molecular system.[29,18,41,43] Notably, the $\omega_{392}$ Mollow triplet is exceptionally more intense than the triplet at the bare resonance $\omega_0$ (Figure 3A). This can be explained by the diminished photon emission at the central resonance due to photon antibunching[17,18] and resonant coupling at $\omega_0$, which suppresses the multi-photon processes involved in the Mollow triplet evolution.[21] In addition, the intense scattering field at P − ($N$ =1), with a mode frequency close to the maximal emission (~ 400 nm) (Figure 2A), favors formation of the $\omega_{392}$ Mollow triplet.[28] This also accounts for the absence of the $\omega_{392}$ Mollow triplet in the $Ni_2$ system, for which the maximal emissions at 380 nm energetically mismatches the $\omega_{392}$ mode.[29]

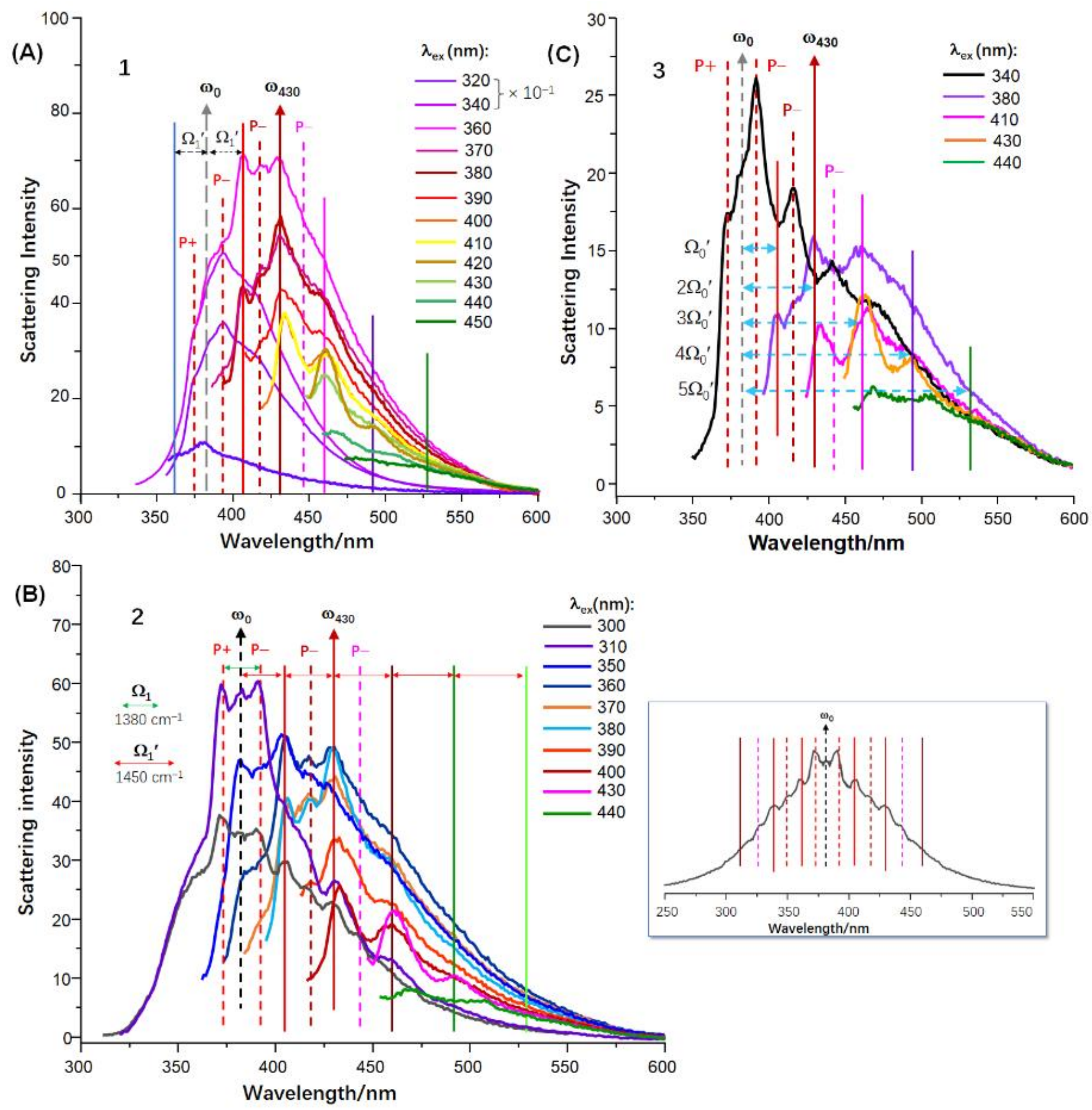

**Figure 4. Quantization and phase transition of the scattering field of the $Mo_2$ complexes 1, 2 and 3.** (A) Spectral variations of **1** with stepwise decrease of the excitation energy, showing the transformation of the R states (dashed vertical lines) to the M states (solid vertical lines) through phase transition. Note that the intensity of the spectra excited at 320 nm and 340 nm is lowered by a factor of ten based on the experimental measurements. (B) Evolution of the Rabi (R) and Mollow (M) transitions for $N$ = 1–3 by stepwise lowering the excitation energy. Excitation at 310–360 nm induces phase-controlled population transfer between the Rabi and Mollow manifolds. Additional Mollow (M) sidebands arise from collective sideband excitation across the JC and Mollow states. The inset shows the simulated full resonance spectrum based on the spectrum excited at 300 nm. (C)

Featured fluorescence spectra for **3**, showing the Mollow sidebands ($\omega_0 - N\Omega_1'$) resulting from continuously lowering the excitation energy ($\lambda_{ex} \geq 380$ nm).

Upon further decreasing the excitation energy ($\lambda_{ex} \geq 360$ nm), complex **1** exhibits the fluorescence spectra showing alternatively distributed Rabi-doublet (R) and Mollow-triplet (M) peaks corresponding to collective coupling of different orders (Figure 4A, Table S2). Adjacent features are separated by approximately 700 $cm^{-1}$, corresponding to $\Omega_1/2$ and indicating successive π-phase shifts between the neighboring dressed states. Excitation at 360 nm produces two peaks at 404 nm and 430 nm, centered at 417 nm and separated by $\Omega_1'$, showing a vacuum Rabi splitting of the $\omega_{417}$ mode. This $\omega_{417}$ nm mode represents a coherent state resulting from a blue (or red) π-phase shift ($\Omega_1'/2$) from the $\omega_{404}$ (or $\omega_{430}$) mode, corresponding to the first or second sideband of the dressed ensemble ($N$ = 2).[29,44] Therefore, this feature. produced by excitation at 370 nm (Figure 4A), demonstrates sideband excitation via a single-photon process governed by photon antibunching.[9,17,45] The resulting P+ (404 nm) and P− (430 nm) branches coincide with the blue sideband and the central line of the $\omega_{430}$ Mollow triplet.[29] This $\omega_{430}$ Mollow triplet spectrum reaches its maximum intensity under resonance excitation ($\lambda_{ex}$ = 380 nm). As the excitation energy is further decreased to 430 nm, additional Mollow sidebands appear sequentially at $\omega_0 - N\Omega_1'$ ($N$ = 4–5), with peaks observed at 490 nm and 528 nm (Figure 4A). The sideband transitions are in quantitative agreement with those observed in the $Ni_2$ system.[29]

$Mo_2(DAniF)_4$ (**2**) produces the resonance fluorescence nearly identical to that observed for its $Ni_2$ analogue, $Ni_2(DAniF)_4$.[29] When excited at 300 nm, the spectrum exhibits a staircase structure for the low-energy polaritonic transitions (Figure 4B, Table S3),[29] with the neighboring steps separated by ~ 700 $cm^{-1}$. From this spectrum, a complete polaritonic spectral profile can be constructed by symmetry generation for the $M_2$ (M = Mo, Ni) molecular quantum systems, yielding transition energies that scale as $\omega_0 \pm N\sqrt{N}\Omega_1$ for the Rabi manifolds and $\omega_0 \pm N\Omega_1'$ for the Mollow manifolds (Figure 4B, inset). The consistency of these behaviors across chemically distinct $Mo_2$ and $Ni_2$ systems demonstrates that the transformation between Rabi splitting and Mollow triplet states is governed by the dynamics of the collective coupling in the JC molecular systems. The spectra obtained with $\lambda_{ex} \geq 380$ nm are very similar to those in system **1**, confirming that the optical responses under non-resonant excitations are dominated by the $M_2$ unit, regardless of their electronic structures that depend on their ligand environments.

Exciting **3** at 340 nm allows the three low-energy Rabi branches to be observed at approximately 392 nm, 413 nm, and 443 nm (see Figure 4C, black trace), resulting from single and collective coupling. These peaks are assigned to the P− polaritons for $N$ = 1, 2, and 3, respectively. The transition energies are in excellent agreement with the values predicted from $[\omega_0 - (N\sqrt{N}\Omega_1)/2]$, while the red Mollow triplet states ($\omega_0 - N\Omega_1'$) correspond to the emission valleys. Resonant excitation at 380 nm produces a spectrum showing three peaks at 404 nm, 430 nm and 458 nm with transitions that correspond to those of the $\omega_{430}$ Mollow triplet observed for **1**, **2** (ref. 28) (Table S4) and $Ni_2$.[29] However, the spectral lineshape significantly deviates from a Mollow triplet profile (Figure 4C, purple trace). The distortion of the $\omega_{430}$ Mollow triplet for **3** may be caused by its asymmetrical molecular structure, which prevents better alignment of the dipole moment of the molecules for collective coupling. There is evident spectral agreement between complexes **1**, **2** and **3** when they are excited at 420 – 430 nm; their spectra show three characteristic Mollow-triplet sidebands at 458 nm, 492 nm and 530 nm, with gradually decreased intensity (Table S4). The spectral profile changes under excitation of 440 nm, showing two additional peaks of similar intensities at approximately 470 nm and 507 nm (Figure 4C). Similar spectra are also observed by excitation at the same wavelength for **1** (Figure 4A) and **2** (Figure 4B). These two peaks are centered at 488 nm, close to the 490 nm sideband, and are separated by 1550 $cm^{-1}$. It is possible that this feature results from the splitting of the $\omega_{490}$ mode via a single-photon coupling process through polariton excitation pathway.[42,45] This excitation-induced transformation from the M state to the R state can be attributed to the low scattering intensity around the excitation wavelength. This is in contrast to the R to M state transition at 380 nm, which is caused by the intense emission under resonant excitation. Nevertheless, the steady-state fluorescence spectra for the $Mo_2$ systems (**1**, **2** and **3**) demonstrate that the optical responses of the dimetal complexes is governed by the $Mo_2$ core that acts as a diatomic optical resonator.

## 3. Discussion and Outlook

Sixty years after the discovery of metal–metal quadruple bonds by F. A. Cotton, this study reveals that the $Mo_2$ unit in quadruply bonded dimolybdenum complexes functions not only as a chemically well-defined structural motif, but also as an intrinsic molecular resonator capable of confining and quantizing visible light. Spectroscopic evidence demonstrates that the two Mo atoms form an atomistic cavity in which the scattering field is strongly enhanced and discretized, enabling coherent light–matter interactions to emerge under ambient conditions. In this sense, the $Mo_2$ unit represents a built-in optical resonator that does not rely on external cavities, nanofabrication, or cryogenic environments. The results show that the intermetallic charge-transfer transition within the $Mo_2$ unit is intrinsically and resonantly coupled to the local scattering field generated by the diatomic metal core. This coupling gives rise to characteristic quantum-optical phenomena, including vacuum Rabi splitting, Mollow triplets, and collective coupling behavior, which are conventionally associated with atomic or solid-state cavity QED systems. Importantly, these effects are observed under weak, non-resonant excitation and are reproducible across chemically distinct $Mo_2$ complexes, indicating that they originate from the diatomic $Mo_2$ resonator itself rather than from ligand-specific electronic transitions.

From a molecular-science perspective, these findings suggest that traditional descriptions of metal–metal bonding based solely on electronic structure and bond order become incomplete when strong coupling to quantized electromagnetic fields is present. In the $Mo_2$ system, electronic and photonic degrees of freedom are inseparably intertwined, giving rise to hybrid exciton–polaritonic states that govern the optical response of the molecule. The $Mo_2$ unit thus behaves as an effective molecular quantum emitter–resonator, in which light confinement and collective interactions redefine the nature of metal–metal bonding under optical excitation. The close correspondence between the quantum-optical behavior of $Mo_2$ and that previously reported for $Ni_2$ systems, despite their markedly different bonding characteristics, establishes the diatomic $M_2$ (M = Ni, Mo) motif as a general molecular platform for studying cavity-free quantum electrodynamics. This universality underscores that the essential requirement for field quantization in these systems is the diatomic metal geometry and sub-angstrom metal–metal separation, rather than a specific electronic configuration or ligand environment. As such, quadruply bonded $Mo_2$ complexes provide a chemically robust and conceptually transparent system for extending quantum-optical concepts into molecular science.

Looking forward, the ability of single molecules to trap and quantize light in free space opens new directions for understanding and controlling chemical behavior through light–matter interactions. Further experimental and theoretical efforts aimed at characterizing the scattering fields and their quantum states will deepen insight into the interplay between molecular structure and quantized electromagnetic environments. Beyond fundamental interest, these advances may ultimately enable new strategies for manipulating chemical reactivity and developing molecular-scale quantum technologies.

## 4. Experimental Section

*Chemicals*

$Mo_2$ complexes $Mo_2(O_2CCH_3)_4$ (**1**),[46] $Mo_2(DAniF)_4$ (**2**)[35] (DAniF = *N*, *N*′-di(*p*-anisyl)formamidiante), and $Mo_2(DAniF)_3(O_2CC_6H_5)$ (**3**)[47] were prepared using the published methods and characterized by $^1H$ NMR spectra.

*Fluorescence spectroscopy*

Steady-state fluorescence spectra were recorded on a Shimadzu RF-5301PC

spectrometer equipped with a 150 W xenon lamp as the excitation source. Excitation wavelengths were selected by the instrument's monochromator with a bandwidth of 5 nm. All measurements were performed at room temperature in dichloromethane (DCM) solutions ($5\times10^{-6}$ mol·L$^{-1}$).

**Supporting Information**

Supporting Information is available from the Wiley Online Library or from the author.

**Acknowledgments**

We acknowledge the primary financial support from the National Natural Science Foundation of China (22171107, 21971088, 21371074), Natural Science Foundation of Guangdong Province (2018A030313894), Jinan University, and the Fundamental Research Funds for the Central Universities.

**Conflict of Interest**

The authors declare no conflict of interest.

**Data Availability Statement**

All data are available in the manuscript or the supplementary materials. The spectra raw data have been deposited at Mendeley (Mendeley Data: doi: 10.17632/mxbg5b62hp.3 and are publicly available as of the date of publication. Additional information about the data will be made available from the corresponding author upon reasonable request.

Switching at the Single-Photon Level. *Phys. Rev. Lett.* **2021**, *127*, 133603.

[15] W. E. Moerner, M. Orrit, Illuminating Single Molecules in Condensed Matter. *Science*, **1999**, *283*, 1670–1676.

[16] M. Orrit, T. Ha, V. Sandoghdar,. Single-molecule optical spectroscopy. *Chem. Soc. Rev.*, **2014**, *43*, 973–976.

[17] H. J. Kimble, M. Dagenais, L. Mandel, Photon antibunching in resonance fluorescence. *Phys. Rev. Lett.* **1977**, *39*, 691–695.

[18] T. Basché, W. E. Moerner, M. Orrit, H. Talon, Photon Antibunching in the Fluorescence of a Single Dye Molecule Trapped in a Solid. *Phys. Rev. Lett.* **1992**, *69*, 1516–1519.

[19] D. F. Walls, Squeezed states of light. *Nature*, **1983,** *306*, 141–146.

[20] X. L. Chu, S. Götzinger, V. Sandoghdar, A single molecule as a high-fidelity photon gun for producing intensity-squeezed light. *Nature Photonics,* **2017**, *11*, 58–62.

[21] B. R. Mollow, Power spectrum of light scattered by two-level systems. *Phys. Rev.* **1969**, *188*, 1969–1975.

[22] J. C. López Carreño, E. del Valle, F. P. Laussy, Photon correlations from the Mollow triplet. *Laser Photonics Rev.* **2017***, 11*, 170009.

[23] D. J. Tibben, G. O. Bonin, I. Cho, G. Lakhwani, J. Hutchison, D. E. Gómez, Molecular energy transfer under the strong light–matter interaction regime. *Chem. Rev.,* **2023***, 123*, 8044–8068.

[24] C R. Chikkaraddy, B. de Nijs, F. Benz, S. J. Barrow, O. A. Scherman, E. Rosta, A. Demetriadou, P. Fox, O. Hess, J. J. Baumberg. Single-molecule strong coupling at room temperature in plasmonic nanocavities. *Nature* **2016**, *535*, 127–130.

[25] F. Benz, M. K. Schmidt, A. Dreismann, R. Chikkaraddy, Y. Zhang, A. Demetriadou, C. Carnegie, H. Ohadi, B. de Nijs, R. Esteban, J. Aizpurua, J. J. Baumberg, Single-molecule optomechanics in "picocavities". *Science* **2016**, *354*, 726–729.

[26] K. J. Savage, M. M. Hawkeye, R. Esteban, A. G. Borisov, J.r Aizpurua, J. J. Baumberg. Revealing the quantum regime in tunnelling plasmonics. *Nature* **2012***, 491*, 574–577.

[27] R. Esteban, A. G. Borisov, P. Nordlander, J. Aizpurua, Bridging quantum and classical plasmonics with a quantum-corrected model. *Nat. Commun.,* **2012**, *3*, 825.

[28] Y. N. Tan, M. Meng, Z. C. He, T. Cheng, J.-B. Luo, X.-D. Wang, Z. H. Zhong, J. Zhou, G. Y.